\RequirePackage{fix-cm}
\documentclass[12pt]{article}
\usepackage[fontsize=13pt]{scrextend}
\usepackage{setspace,graphicx,epstopdf,amsmath,amsfonts,amssymb,amsthm, lscape, flafter, hyperref,float, caption, tabularx, marginnote,datetime,enumitem,subfigure,rotating,fancyvrb, mathtools, booktabs, changepage, setspace, placeins, threeparttable, ragged2e, stmaryrd}
\usepackage[longnamesfirst]{natbib}
\usepackage[export]{adjustbox}
\newcolumntype{Y}{>{\raggedleft\arraybackslash}X}
\usdate
\usepackage[a4paper,bindingoffset=0.2in,%
            left=0.8in,right=0.8in,top=1in,bottom=0.8in,%
            footskip=.35in]{geometry}

\usepackage{indentfirst} 
\usepackage{endnotes}    

\begin{document}

\setlist{noitemsep}  
\onehalfspacing      
\renewcommand{\footnote}{\endnote}  

\author{\large{Mykola Pinchuk}\thanks{\rm Simon Business School, University of Rochester. Email: Mykola.Pinchuk@ur.rochester.edu.}}

\title{\bf Bitcoin Does Not Hedge Inflation}

\date{30 May 2021}  

\maketitle
\thispagestyle{empty}

\bigskip

\normalsize

\vspace{1cm}

\centerline{\bf Abstract}

\vspace{0.5cm}

\begin{onehalfspace}  
  \noindent This paper examines the response of major cryptocurrencies to macroeconomic news announcements (MNA). While other cryptocurrencies exhibit no reaction to major MNA, Bitcoin responds negatively to inflation surprise. Price of Bitcoin decreases by 24 bps in response to a 1 standard deviation inflationary surprise. This reaction is inconsistent with widely-held beliefs of practitioners that Bitcoin can hedge inflation. I do not find support for the hypothesis that the negative response of Bitcoin to inflation is due to its negative exposure to interest rates. Instead, I find support for the hypothesis that Bitcoin is strongly affected by the shift in consumption-savings decisions, driven by the rise in inflation. Consistent with this view, Bitcoin has negative exposure to a proxy for the consumption-savings ratio.
\end{onehalfspace}
\medskip

\clearpage
\setstretch{1.5}

\section{Introduction} \label{sec:Model}
Since the introduction of Bitcoin in 2009, cryptocurrencies have been the most rapidly growing asset class. Between January 2009 and May 2021, the market capitalization of the cryptocurrency market surged from 0 to 1.4 trillion dollars. Such a rapid increase has been puzzling for academics and practitioners alike. Unlike traditional asset classes, cryptocurrencies do not produce cash flows in the future, which renders their valuation problematic. Unlike traditional currencies, cryptocurrencies are not recognized and enforced as a valid means of payment by the state. So it is not surprising that both academic researchers and industry practitioners have been struggling to explain what gives value to cryptocurrencies.
\paragraph{}
While cryptocurrencies are viewed as a single asset class, it is possible that different cryptocurrencies derive their value from different channels. Ethereum as well as cryptocurrencies that compete with it (e.g., Cardano) or use its ecosystem (e.g., Uniswap) could have various applications, such as low-cost payment systems, decentralized finance, decentralized autonomous organizations, or prediction markets. However, Bitcoin and similar cryptocurrencies account for most of the cryptocurrency market capitalization. Since Bitcoin is not useful for the aforementioned applications, its proponents argue that it derives its value from being "digital gold". According to this view, Bitcoin can displace fiat money by being a better store of value. Unlike traditional currencies, Bitcoin is not controlled by the central government, which can make it a better store of value. Since the supply of Bitcoin grows at a steady and very low predetermined rate, it is supposed to be a hedge against inflation. Being a hedge for inflation is usually cited as the main advantage of Bitcoin. Since owning cash is much cheaper and more convenient than holding Bitcoin, the inflation-hedging benefits of Bitcoin must be large to justify its use. 
\paragraph{}
While theoretically, we can expect Bitcoin to be good inflation hedge, ultimately, this is an empirical question. In this paper, I attempt to empirically explore the exposure of cryptocurrencies to various macroeconomic variables, including inflation. There are two approaches to estimating the exposure of financial assets to macroeconomic variables. The most commonly used approach is estimating the correlation between the returns of the asset and the time series of some macroeconomic variable. Since most macroeconomic variables are collected and updated at a low frequency, this usually means running regressions at a monthly or quarterly frequency. An alternative, less commonly used approach is to look at the response of the price of the asset to an unexpected component in a macroeconomic news announcement (MNA). This approach requires using intraday pricing data and has a number of advantages, including accounting for market expectations, higher signal-to-noise ratio, increased statistical power by combining multiple MNA of the same type and the ability to give causal interpretation to the results. This paper utilizes the second approach, using 30-minute window around MNA.
\paragraph{}
I explore 6 major cryptocurrencies (Bitcoin, Ethereum, Ripple, Litecoin, Cardano, and Bitcoin Cash) and find that except Bitcoin, they have no significant loadings on macroeconomic variables. While Bitcoin has mostly insignificant exposure to macroeconomic variables, proxying for economic growth, it has large negative exposure to inflation surprise. In response to 1 standard deviation inflationary surprise, Bitcoin price decreases by 24 bps (t-statistics 2.44). This result is puzzling since theoretically Bitcoin is supposed to serve as an inflation hedge. In other words, we could expect that Bitcoin will have positive exposure to inflation surprise. The negative loading of Bitcoin on inflation news casts doubt on its ability to be a better store of value than traditional fiat currency. 
\paragraph{}
In the second part of the paper, I explore two non-mutually exclusive explanations of the negative exposure of Bitcoin to inflation. First, it is likely that higher inflation leads to an increase in expected future interest rates because of expected monetary tightening by the Fed. A higher interest rate could decrease the price of Bitcoin due to a variety of reasons. For example, it could mean discounting whatever benefits Bitcoin provides at a higher rate. Alternatively, higher rates will make the investment into bonds more attractive, raising the opportunity cost of holding Bitcoin. The second hypothesis entails that Bitcoin has negative exposure to inflation due to the effect of inflation on the consumption-savings decisions of investors. Large inflationary surprise increases inflationary expectations, prompting people to consume more and save less. Lower savings lead to lower net inflows to financial assets, including Bitcoin, causing a drop in Bitcoin prices.
\paragraph{}
I find little support for the hypothesis, according to which interest rate movement explains the negative response of Bitcoin to inflation surprises. While interest rate responds positively to inflation surprise, there is no evidence of negative comovement between Bitcoin price and interest rates during the inflationary news window. If anything, this comovement is positive, but not statistically significant (t-statistic of 0.9). Overall, the increasing interest rates can not explain negative response of Bitcoin to rising inflation.
\paragraph{}
I find evidence, supporting the hypothesis, according to which the negative response of Bitcoin to inflation is driven by the effect of inflation on the consumption-savings decisions of investors. I argue that due to the timing of announcements of macroeconomic news, aggregate consumption news surprise contains two components: surprise in the level of consumption and surprise in the consumption-savings ratio. The fact that Bitcoin has negative exposure to consumption news surprise and zero exposure to all other growth-related news surprises implies that it has negative loading on the consumption-savings ratio. These results are consistent with anecdotal evidence that Bitcoin surges coincide with periods of excess savings. 
\paragraph{}
This paper is related to two strands of research in asset pricing and macrofinance. First, it contributes to the growing cryptocurrency literature. Schilling and Uhlig (2019) introduce a model of cryptocurrency and explain possible reasons why cryptocurrency could have value. Liu and Tsyvinski (2019, 2021) study key empirical patterns in cryptocurrency returns. The closest paper to this one is Liu and Tsyvinski (2021), which briefly explores the loadings of cryptocurrencies on macroeconomic factors and finds no significant results. My paper differs in using event study methodology to estimate loadings of cryptocurrencies on macroeconomic variables. As I explain in Section 3, this methodology has a number of advantages, which allow me to get cleaner estimates of the exposure of cryptocurrencies to macroeconomic variables. Second, it contributes to empirical macrofinance literature, which attempts to connect asset prices to changes in the macroeconomic environment. Schwert (1981), Pearce and Roley (1985), Roley and McQueen (1993), Flannery and Protopapadakis (2002), Boyd, Hu and Jagannathan (2005) are examples of the papers, using event studies to estimate the response of financial assets to macroeconomic news. This paper uses a similar event studies methodology to Pinchuk (2021), which studies the response of the equity market to macroeconomic news.
\paragraph{}
Overall, these results could be of interest to both academic researchers and industry practitioners. The fact that Bitcoin loses its value in response to an unexpected rise in inflation means that investors should not rely on Bitcoin to hedge inflation risk. More generally, this fact, together with the negative exposure of Bitcoin to the consumption-savings ratio, points to an explanation of Bitcoin pricing by fund flows, potentially unrelated to fundamental forces. This view is consistent with behavioral explanations of surges in Bitcoin prices.

\section{Data} 
I obtain the data on MNA from Bloomberg. For each type of MNA, Bloomberg provides data on the exact timing of the release, announced value of macroeconomic variables and its previous value. In order to avoid data snooping concerns, I focus on the subset of MNA, which were shown to be important to equity and bond returns (Savor and Wilson 2013, Hu, Pang, Wang, Zhu 2019, Ernst, Gilbert and Hrdlicka 2020, Pinchuk 2021). This subset consists of the following news:
\begin{itemize}
    \item {GDP.}
    \item {Personal consumption expenditure (PCE).}
    \item {Non-farm payrolls (NFP).}
    \item {Unemployment rate.}
    \item {ISM Manufacturing.}
    \item {Construction spending.}
    \item {Retail sales.}
    \item {Consumer Price Index (CPI).}
    \item {Producer Price Index (PPI).}
\end{itemize}

Table 1 describes the sample of MNA. I divide them into two groups, depending on the nature of announced macroeconomic variable. While CPI and PPI news are inflation news, all other MNA can be seen as proxies for the news about economic growth. The sample spans the period between 1 January 2013 and 12 May 2021 and includes 99-101 observations for each type of MNA. While all these MNA are released at monthly frequency, GDP and PCE news are the only types of MNA, which contain data for the previous quarter. After the initial release of the data on GDP and PCE in the previous quarter, US Bureau of Economic Analysis releases updates at the end of the two following months. Since updates are usually very close to initial announcements, in the next sections of the paper I will use only initial announcements of GDP and PCE news. All MNA are released at 8:30 or 10:00am. 
\paragraph{} 
As Brunnermeier, Fahri, Koijen, Krishnamurthy, Ludvigson, Lustig, Nagel and Piazzesi (2021) point out, using proxies for investors' expectation is very important to study asset prices. I use surveys of investors` forecasts, conducted by Bloomberg before each MNA. Within two weeks prior to MNA release, Bloomberg surveys more than 100 professional market participants in order to elicit their forecasts of upcoming MNA. Bloomberg allows respondents to update their forecasts up to 2 hours before MNA. Hence it seems reasonable to assume that these forecasts are timely proxies for expectations of investors about upcoming release. I construct MNA surprise as the difference between actually announced and mean expected values of macroeconomic variables. Table 2 describes surprises for all 9 types of MNA. Surprises for each MNA except Construction spending are centered around zero, suggesting absence of bias in survey forecasts.
\paragraph{} 
I merge the data of MNA with prices of 6 major cryptocurrencies. Beginning of the sample period differs across cryptocurrencies. Bitcoin has the longest sample, covering the period between March 2013 and May 2021. For each cryptocurrency, I use pricing data, aggregated at 1-minute frequency. These data are available 24 hours per day, which is very convenient for MNA event studies, given that most MNA are released at 8:30 in the morning. 
\paragraph{} 
Table 3 documents summary statistics of daily cryptocurrency returns over the sample. Mean returns of these cryptocurrencies are 10-34 bps and daily Sharpe Ratio is 20-84\%. Table 4 presents summary statistics of Bitcoin returns during 30-minutes window, which starts 10 minutes before and ends 20 minutes after MNA. For all these MNA, Bitcoin returns are centered around zero and do not exhibit strong evidence of announcement premium. To proxy for changes in interest rate, I use intraday futures prices for 5-year Treasury notes. The sample includes period between 1 January 2013 and August 2020.

\section{Using event study to estimate exposure to macroeconomic variables} 

When trying to test macrofinance models, researchers usually estimate exposures of financial assets to macroeconomic variables using low-frequency regressions. In other words, they use regression of time series of return of a given asset on time series of levels or differences in some macroeconomic variable. Usually such regressions use monthly, quarterly or annual frequency. While asset pricing literature has been using this approach for at least 40 years, most of the time it either results in lack of any statistically significant results or finding correlation between some asset returns and every important macroeconomic variable. In other words, even when we find some correlation, it is usually impossible to trace it to a particular macroeconomic variable. Moreover, due to a low frequency, these results are often vulnerable to reverse causality and hence lack causal interpretation. 
\paragraph{} 
Brunnermeier, Fahri, Koijen, Krishnamurthy, Ludvigson, Lustig, Nagel and Piazzesi (2021) highlight the importance of accounting for expectations of investors, which is very problematic within traditional approach. By definition, price of an asset is a function of the information set of investors. Hence any price movement should be driven by the arrival of new information. When using traditional low-frequency regressions, researchers usually implicitly assume that macroeconomic variable is a martingale (when using first differences) or that it follows some AR process (when using residuals from some time-series model). Extensive literature on forecasting of macroeconomic variables suggests that time-series models perform poorly compared to survey forecasts (Ang, Bekaert and Wei 2007). To the best of my knowledge, there is no source of forecasts of macroeconomic variables at monthly frequency, made immediately before the beginning of forecast period. 
\paragraph{}
Event studies are surprisingly underutilized alternative. Using event study to estimate exposures of financial asset to macroeconomic variables means measuring returns during the short window around MNA and regressing these returns on announcement surprise. This approach requires high-frequency pricing data as well as survey forecasts of MNA and has a number of advantages:
\begin{enumerate}
    \item {Measuring surprise in MNA as opposed to change from the previous period.}
    \item {Higher signal-to-noise ratio.}
    \item {Results are not subject to reverse causality.}
    \item {It is possible to distinguish important $MNA_1$ from unimportant $MNA_2$, correlated with $MNA_1$.}
    \item {Combining multiple MNA of a similar type allows to effectively increase sample size and gain larger statistical power.}  
\end{enumerate}
\paragraph{}
Increased statistical power is especially important given that cryptocurrencies have only 4-8 years of reliable pricing data. Overall, using event study methodology to estimate exposure to macroeconomic news allows to improve accuracy and statistical power. Moreover, it shifts interpretation of the results from mere correlations towards causal effect of MNA on asset prices.
\paragraph{}
I estimate all regressions using 30-minutes window around MNA. The window starts 10 minutes before MNA and ends 20 minutes after it and is the most commonly used window in intraday event study literature. In other words, I estimate regression (1):
\begin{equation}
    R_t = a+b*Surprise_t+\epsilon_t.
\end{equation}
$Surprise_t$ is computed as a difference between actually announced value of macroeconomic variable and the mean forecast from Bloomberg surveys. Since most Bloomberg survey responses as of the time of MNA are less than 5 days old, they give timely proxy for investors' expectations. 

\section{Results} 

This section of the paper studies response of cryptocurrencies to MNA surprises. First, I divide MNAs into 2 groups: news about economic growth and news about inflation. Economic growth news include GDP, Personal consumption, Non-farm payroll, Unemployment, PMI, Construction spending and Retail sales. Inflation news include CPI and PPI news. Combining multiple MNA into a group allows to effectively increase sample size and gain extra statistical power. 
\paragraph{}
Table 5 reports the response of the prices of major cryptocurrencies to MNA about economic growth. No cryptocurrency exhibits significant reaction to these news. These responses are not significant either statistically (t-statistic below 0.5) or economically (less than 2 bps per 1 standard deviation surprise). Since sample size is between 268 and 680 observations for different cryptocurrencies, low statistical power can not explain these results.
\paragraph{}
Table 6 presents the response of cryptocurrency prices to MNA about inflation. Since inflation news include only 2 MNA (CPI and PPI), sample size is smaller, which could lead to some concerns about lack of statistical power. Ethereum, Litecoin and Cardano do not have significant exposure to inflation news. Ripple decreases by 11 bps in response to inflation surprise, which is not statistically significant (t-statistic of 1.0). Bitcoin and Bitcoin Cash exhibit large negative loading on inflation, losing 24 bps and 18 bps respectively in response to 1 standard deviation of inflation surprise. Negative reaction of Bitcoin to inflation surprise is highly statistically significant with t-statistic of 2.44 and explains 2.5\% variation in announcement-window returns of Bitcoin.
\paragraph{}
This result is puzzling, since Bitcoin is supposed to hedge inflation. In other words, theoretically we could expect positive exposure of Bitcoin to inflation surprise. Negative loading of Bitcoin on inflation means that it loses value as inflation accelerates, so investors should not use it as inflation hedge. Most importantly, this result contradicts the view of Bitcoin as alternative means of store of value. Since store of value is seen as a primary function of Bitcoin, this result raises question as to why Bitcoin has value. 

\section{Mechanism} 

This section explores two non-exclusive explanations of negative exposure of Bitcoin to inflation. One possible explanation of the negative exposure of Bitcoin to inflation relies on a change in interest rates. The dual mandate of the Fed means that it sets short-term interest rates to achieve stable prices and full employment. Thus inflationary surprise will increase probability of inflation exceeding Fed's target and will make rate hike in the future more likely. According to this hypothesis, higher inflation raises risk-free rate, which could put downward pressure on cryptocurrencies. One possible reason for negative comovement between cryptocurrencies and interest rates is increased opportunity cost of holding cryptocurrencies due to higher yield on government bonds. This view is similar to so-called Fed model or various models of reaching for yield. Another possible reason for negative comovement between cryptocurrencies and rates is higher funding cost, resulting from higher rates. Large fraction of trading in cryptocurrencies takes place on unregulated exchanges, where investors employ very high leverage. Therefore it is possible that cryptocurrencies are vulnerable to adverse changes in funding conditions. As interest rates increase, scarcer liquidity could lead to sell-offs in cryptocurrency market.
\paragraph{}
Table 8 shows response of interest rate to MNA surprise among 2 groups of news. 5-year treasury rates increase in response to both faster economic growth (column 1) and higher inflation (column 2). These responses are highly statistically significant with t-statistic of 10 and 7 respectively. So far these results are consistent with the hypothesis. 
\paragraph{}
Table 9 reports results of the regression of Bitcoin returns on the change in interest rates during MNA window. There is no significant relation during either growth or inflation news. If anything, the relation between Bitcoin price and interest rates during inflation news is positive, though not statistically significant. This result is inconsistent with the hypothesis that negative loading of Bitcoin on inflation reflects the effect of interest rates.
\paragraph{}
Another explanation of the negative exposure of Bitcoin to inflation relies upon a change in consumption-savings decisions of investors. Positive inflationary surprise means that inflation in the previous month was higher than expected. According to multiple-equilibria view of inflation, inflation is usually "anchored" at a target rate. Any significant deviation from this target makes "unanchoring" more likely. Hence large inflationary surprise will increase inflationary expectations of investors. Higher inflationary expectations imply that current consumption becomes cheaper, which leads to an increase in consumption and a decrease in savings. This could significantly affect high-risk assets like cryptocurrencies.
\paragraph{}
Ideally, to test this hypothesis, we need a proxy for consumption/savings ratio. Mismatch in timing of the release of consumption and economic growth data allows me to use personal consumption expenditure data to proxy for this ratio. Data on personal consumption and savings for month t are released at the end of month t+1. By that time market learns realizations of several macroeconomic variables, proxying for economic growth and income (employment, unemploymnet, industrial production and construction spending), so investors have reasonably good expectations about personal income in month t before the release of these data. Thus when data on income, consumption and savings are released at the end of month t+1, it mostly resolves uncertainty on the distribution of this income between consumption and savings. Since Bloomberg collects survey forecasts only on personal consumption, I use surprise in personal consumption to proxy for both consumption level news and consumption/savings rate news. Hence the sensitivity of returns of an asset to personal consumption news surprise is a sum of its sensitivity to personal consumption and its sensitivity to consumption/savings ratio.
\paragraph{}
Table 7 shows that, unlike most financial assets, Bitcoin has zero loading on various proxies for income growth and economic growth, excluding PCE. Thus it is reasonable to assume that its sensitivity to PCE growth is zero. This implies that its negative loading on PCE news surprise reflects its negative exposure to consumption/savings ratio. Then the interpretation of these results is that Bitcoin depreciates in response to a decrease in savings rate. It is consistent with the intuition, according to which people invest large fraction of excess savings in high-risk assets such as cryptocurrencies. 
\paragraph{}
While supported by the data, this hypothesis faces some theoretical and empirical challenges. Theoretical literature in asset pricing usually views consumption-saving decisions of investors as a function of changes in fundamental quantities, such as risk aversion, long-run risk or probability of rare disaster. Unlike this literature, my hypothesis treats a part of variation in consumption-savings ratio as exogenous. Spike in excess savings during recent Covid-19 period is one of examples of such exogenous variation, when increase in savings was likely driven by the factors other than consumption-savings decisions of investors. Another open question is whether cryptocurrencies are more exposed to consumption-savings ratio surprises than other asset classes. While it is possible that positive response to savings rate shock holds for most financial assets, I can not test this idea using these data. The reason is that most assets have non-zero loadings on economic growth surprises, which makes a loading on consumption/savings ratio unidentified.  

\section{Conclusion} 

This paper studies the response of 6 major cryptocurrencies to Macroeconomic News Announcements (MNA). Using high-frequency data and event study methodology allows to better estimate exposures of financial assets to macroeconomic variables. I find that Bitcoin is the only cryptocurrency, which has significant exposure to some MNA. While it has no exposure to growth-related MNA, Bitcoin has significant negative exposure to inflation shocks. Its price decreases by 24 bps in response to 1 standard deviation inflationary surprise. This response implies that Bitcoin suffers from high inflation and is the opposite of the conventional view of Bitcoin as inflation hedge.
\paragraph{}
I explore two possible explanations for such negative exposure of Bitcoin to inflation. I do not find evidence in favor of the hypothesis that the negative exposure of Bitcoin to interest rate changes drives this negative response. I find support for the hypothesis that Bitcoin suffers from inflation since higher inflation increases current consumption and decreases savings. Consistent with this view, Bitcoin has negative exposure to a proxy for the consumption-savings ratio. This could imply that Bitcoin is among financial assets, the most sensitive to excess savings.
\paragraph{}
Overall, my results suggest that investors should not use Bitcoin to hedge inflation. More broadly, they raise questions on the feasibility of using Bitcoin as a store of value and imply that large surges in its price are primarily driven by non-fundamental forces.

\pagebreak

\section{References:}
\begin{enumerate}
    \item{Andersen, Torben G., Tim Bollerslev, Francis X. Diebold, and Clara Vega. "Micro effects of macro announcements: Real-time price discovery in foreign exchange." American Economic Review 93, no. 1 (2003): 38-62.}
    \item{Bernanke, Ben and Kuttner, Kenneth (2005), What Explains the Stock Market's Reaction to Federal Reserve Policy?. The Journal of Finance, 60: 1221-1257.}
    \item{Boyd, John H., Jian Hu, and Ravi Jagannathan. "The stock market's reaction to unemployment news: Why bad news is usually good for stocks." The Journal of Finance 60, no. 2 (2005): 649-672.}
    \item{Brunnermeier, Markus, Emmanuel Farhi, Ralph SJ Koijen, Arvind Krishnamurthy, Sydney C. Ludvigson, Hanno Lustig, Stefan Nagel, and Monika Piazzesi. "Perspectives on the Future of Asset Pricing." The review of financial studies (2021).}
    \item{Cong, Lin William, and Zhiguo He. "Blockchain disruption and smart contracts." The Review of Financial Studies 32, no. 5 (2019): 1754-1797.}
    \item{Ernst, Rory, Thomas Gilbert, and Christopher M. Hrdlicka. "More than 100\% of the equity premium: How much is really earned on macroeconomic announcement days?." Available at SSRN 3469703 (2019).}
    \item{Flannery, Mark J., and Aris A. Protopapadakis. "Macroeconomic factors do influence aggregate stock returns." The review of financial studies 15, no. 3 (2002): 751-782.}
    \item{Hanson, Samuel G., and Jeremy C. Stein. "Monetary policy and long-term real rates." Journal of Financial Economics 115, no. 3 (2015): 429-448.}
    \item{Hu, Xing, Jun Pan, Jiang Wang, and Haoxiang Zhu. Premium for heightened uncertainty: Solving the fomc puzzle. National Bureau of Economic Research, 2019.}
    \item{Kothari, Sagar P., and Jerold B. Warner. "Econometrics of event studies." In Handbook of empirical corporate finance, pp. 3-36. Elsevier, 2007.}
    \item{Liu, Yukun, and Aleh Tsyvinski. "Risks and returns of cryptocurrency." The Review of Financial Studies 34, no. 6 (2021): 2689-2727.}
    \item{Liu, Yukun, Aleh Tsyvinski, and Xi Wu. Common risk factors in cryptocurrency. No. w25882. National Bureau of Economic Research, 2019.}
    \item{McQueen, Grant, and V. Vance Roley. "Stock prices, news, and business conditions." The review of financial studies 6, no. 3 (1993): 683-707.}
    \item{Pearce, Douglas K., and V. Vance Roley. Stock prices and economic news. No. w1296. National bureau of economic research, 1984.}
    \item{Pinchuk, Mykola. Response of stock market to Macroeconomic News Announcements. Working paper, 2021.}
    \item{Schilling, Linda, and Harald Uhlig. "Some simple bitcoin economics." Journal of Monetary Economics 106 (2019): 16-26.}
    \item{Schwert, G. William. "The adjustment of stock prices to information about inflation." The Journal of Finance 36, no. 1 (1981): 15-29.}
\end{enumerate}

\pagebreak

\section{Appendix} 

\begin{table}[!htbp] \centering 
  \caption{\textbf{Main MNA}} 
  \label{} 
    \begin{flushleft}
    {\medskip\small
 The table reports main details about all MNA, which are relevant for stocks and bonds. N stands for the number of observations in the sample, while Day means business day of the month, when the news is released. I divide MNA into two types: news about economic growth and news about inflation.}
    \medskip
    \end{flushleft}
\begin{tabular}{@{\extracolsep{5pt}} lcccc} 
\\[-1.8ex]\hline 
\hline \\[-1.8ex] 
News & News\_Type & N & Time & Day \\ 
\hline \\[-1.8ex] 
Construction Spending MoM & Growth & $101$ & 10:00 & 1-2 \\ 
ISM Manufacturing & Growth & $101$ & 10:00 & 1-2 \\ 
Change in Nonfarm Payrolls & Growth & $101$ & 08:30 & 1-5 \\ 
Unemployment Rate & Growth & $101$ & 08:30 & 1-5 \\ 
Retail Sales Advance MoM & Growth & $100$ & 08:30 & 9-12 \\ 
PPI MoM & Inflation & $101$ & 08:30 & 9-13 \\ 
CPI MoM & Inflation & $101$ & 08:30 & 7-12 \\ 
GDP Annualized QoQ & Growth & $99$ & 08:30 & 16-21 \\ 
Personal Consumption & Growth & $99$ & 08:30 & 16-21 \\ 
\hline \\[-1.8ex] 
\end{tabular} 
\end{table}

\begin{table}[!htbp] \centering 
  \caption{\textbf{Summary statistics of MNA surprises}} 
  \label{} 
      \begin{flushleft}
    {\medskip\small
 The table reports summary statistics of MNA surprises for all 9 MNA. I report minimum, maximum, mean, median and different percentiles of each MNA surprise. MNA surprise is defined as difference between actually announced and mean expected values of respective macroeconomic variables. The subsample includes all MNA between March 31, 2013 and May 12, 2021. For each variable, surprises are rescaled to have standard deviation of 1 over the whole sample, i.e., 1997-2021.}
    \medskip
    \end{flushleft}
\begin{tabular}{@{\extracolsep{-5pt}} lcccccccccccc} 
\\[-1.8ex]\hline 
\hline \\[-1.8ex] 
News & Min & p1 & p10 & p25 & Median & p75 & p90 & p99 & Max & Mean & SD \\ 
\hline \\[-1.8ex] 
NFP & $-1.11$ & $-0.88$ & $-0.17$ & $-0.06$ & $0.02$ & $0.08$ & $0.15$ & $2.94$ & $16.44$ & $0.20$ & $1.71$ \\ 
PMI & $-2.45$ & $-2.24$ & $-1$ & $-0.46$ & $0.08$ & $0.70$ & $1.37$ & $2.41$ & $2.86$ & $0.13$ & $0.96$ \\ 
Retail & $-4.50$ & $-2.63$ & $-0.65$ & $-0.41$ & $-0.10$ & $0.20$ & $0.41$ & $4.50$ & $9.50$ & $0.04$ & $1.34$ \\ 
Construction & $-3.17$ & $-2.48$ & $-1.64$ & $-0.92$ & $-0.31$ & $0.20$ & $0.82$ & $4.21$ & $4.50$ & $-0.30$ & $1.13$ \\ 
Unemployment & $-14.61$ & $-3.92$ & $-0.77$ & $-0.26$ & $0$ & $0.26$ & $0.26$ & $0.79$ & $1.54$ & $-0.34$ & $1.62$ \\ 
CPI & $-2.40$ & $-1.63$ & $-0.80$ & $-0.80$ & $0$ & $0$ & $0.80$ & $2.48$ & $4.81$ & $-0.04$ & $0.94$ \\ 
PPI & $-1.94$ & $-1.94$ & $-0.83$ & $-0.24$ & $0$ & $0.49$ & $0.73$ & $1.25$ & $2.19$ & $-0.01$ & $0.69$ \\ 
GDP & $-2.73$ & $-2.33$ & $-0.74$ & $-0.21$ & $0$ & $0.63$ & $1.16$ & $2.36$ & $3.36$ & $0.15$ & $0.92$ \\ 
PCE & $-8.47$ & $-3.24$ & $-0.74$ & $-0.21$ & $0.11$ & $0.64$ & $1.16$ & $2.20$ & $3.81$ & $0.10$ & $1.24$ \\ 
\hline \\[-1.8ex] 
\end{tabular} 
\end{table}

\begin{table}[!htbp] \centering 
  \caption{\textbf{Summary statistics of cryptocurrency returns}} 
  \label{} 
     \begin{flushleft}
    {\medskip\small
 The table reports summary statistics of returns of 6 major cryptocurrencies over the sample between March 31, 2013 and May 12, 2021. The columns 2 and 3 specify beginning and end of return sample for each cryptocurrency. The other columns report summary statistics of cryptocurrency daily returns as well as daily Sharpe Ratio.}
    \medskip
    \end{flushleft}
\begin{tabular}{@{\extracolsep{0pt}} lcccccccccc} 
\\[-1.8ex]\hline 
\hline \\[-1.8ex] 
crypto & start & end & min & p5 & median & p95 & max & mean & SD & SR \\ 
\hline \\[-1.8ex] 
ADA & $2018$ & $2021$ & $-40.015$ & $-9.831$ & $0.086$ & $10.141$ & $25.680$ & $0.141$ & $6.187$ & $0.362$ \\ 
BCH & $2017$ & $2021$ & $-46.764$ & $-10.715$ & $0.129$ & $10.544$ & $75.796$ & $0.097$ & $7.745$ & $0.199$ \\ 
BTC & $2013$ & $2021$ & $-63.665$ & $-6.513$ & $0.173$ & $7.160$ & $34.446$ & $0.212$ & $4.856$ & $0.693$ \\ 
ETH & $2016$ & $2021$ & $-41.825$ & $-8.432$ & $0.269$ & $9.351$ & $38.576$ & $0.304$ & $5.780$ & $0.835$ \\ 
LTC & $2016$ & $2021$ & $-36.862$ & $-9.425$ & $0.207$ & $10.274$ & $62.954$ & $0.268$ & $6.539$ & $0.651$ \\ 
XRP & $2017$ & $2021$ & $-40.704$ & $-9.580$ & $0.086$ & $10.809$ & $77.333$ & $0.340$ & $7.738$ & $0.697$ \\ 
\hline \\[-1.8ex] 
\end{tabular} 
\end{table}

\begin{table}[!htbp] \centering 
  \caption{\textbf{Summary statistics of Bitcoin returns during MNA}} 
  \label{} 
       \begin{flushleft}
    {\medskip\small
 The table reports summary statistics of Bitcoin returns during 30-minutes window, centered around MNA release time. Column 1 specifies MNA type, while the other columns report summary statistics of daily returns.}
    \medskip
    \end{flushleft}
\begin{tabular}{@{\extracolsep{-5pt}} lcccccccccccc} 
\\[-1.8ex]\hline 
\hline \\[-1.8ex] 
News & Min & p1 & p10 & p25 & Median & p75 & p90 & p99 & Max & Mean & SD \\ 
\hline \\[-1.8ex] 
NFP & $-4.70$ & $-2.32$ & $-0.52$ & $-0.23$ & $0$ & $0.20$ & $0.64$ & $1.65$ & $1.94$ & $-0.04$ & $0.76$ \\ 
PMI & $-2.15$ & $-1.86$ & $-0.45$ & $-0.23$ & $-0.01$ & $0.19$ & $0.54$ & $1.67$ & $2.12$ & $0$ & $0.56$ \\ 
Retail & $-3.16$ & $-3.05$ & $-0.62$ & $-0.17$ & $0.01$ & $0.26$ & $0.62$ & $2.03$ & $8.07$ & $0.02$ & $1.09$ \\ 
Construction & $-2.15$ & $-1.87$ & $-0.48$ & $-0.23$ & $-0.01$ & $0.18$ & $0.43$ & $1.67$ & $2.12$ & $-0.02$ & $0.55$ \\ 
Unemployment & $-4.70$ & $-2.32$ & $-0.52$ & $-0.23$ & $0$ & $0.20$ & $0.64$ & $1.65$ & $1.94$ & $-0.04$ & $0.76$ \\ 
CPI & $-3.16$ & $-1.55$ & $-0.62$ & $-0.23$ & $-0.03$ & $0.13$ & $0.45$ & $2.66$ & $10.26$ & $0.05$ & $1.23$ \\ 
PPI & $-3.04$ & $-1.88$ & $-0.59$ & $-0.15$ & $0.03$ & $0.26$ & $0.54$ & $2.23$ & $8.07$ & $0.06$ & $1.03$ \\ 
GDP & $-3.22$ & $-2.48$ & $-0.53$ & $-0.20$ & $0$ & $0.19$ & $0.46$ & $2.24$ & $2.92$ & $-0.02$ & $0.72$ \\ 
PCE & $-3.22$ & $-2.48$ & $-0.53$ & $-0.20$ & $0$ & $0.19$ & $0.46$ & $2.24$ & $2.92$ & $-0.02$ & $0.72$ \\ 
\hline \\[-1.8ex] 
\end{tabular} 
\end{table}

\begin{table}[!htbp] \centering 
  \caption{\textbf{Response of cryptocurrencies to MNA about growth}} 
  \label{}
  \begin{flushleft}
    {\medskip\small
 The table documents response of cryptocurrency prices to MNA surprises. Each column reports estimates from the regression $R_t = a + b Surprise_t + \epsilon_t.$ for different cryptocurrency. Cryptocurrency returns are measured over 30-minutes window around MNA.}
    \medskip
    \end{flushleft}
\begin{tabular}{@{\extracolsep{0pt}}lcccccc} 
\\[-1.8ex]\hline 
\hline \\[-1.8ex] 
 & \multicolumn{6}{c}{\textit{Dependent variable:}} \\ 
\cline{2-7} 
\\[-1.8ex] & \multicolumn{6}{c}{Return} \\ 
\\[-1.8ex] & (1) & (2) & (3) & (4) & (5) & (6)\\ 
\hline \\[-1.8ex] 
 Surprise & -0.001 & -0.011 & 0.001 & 0.008 & 0.004 & 0.015 \\ 
  & [-0.037] & [-0.400] & [0.025] & [0.115] & [0.040] & [0.487] \\ 
  & & & & & & \\ 
 Constant & -0.018 & 0.026 & -0.106 & 0.135 & 0.295$^{**}$ & 0.056 \\ 
  & [-0.636] & [0.617] & [-1.346] & [1.106] & [1.969] & [0.972] \\ 
  & & & & & & \\ 
\hline \\[-1.8ex] 
Coin/token & BTC & ETH & XRP & BCH & LTC & ADA \\ 
Observations & 680 & 432 & 359 & 315 & 366 & 268 \\ 
Adjusted R$^{2}$ & -0.001 & -0.002 & -0.003 & -0.003 & -0.003 & -0.003 \\ 
\hline 
\hline \\[-1.8ex] 
\textit{Note:}  & \multicolumn{6}{r}{$^{*}$p$<$0.1; $^{**}$p$<$0.05; $^{***}$p$<$0.01} \\ 
\end{tabular} 
\end{table}

\begin{table}[!htbp] \centering 
  \caption{\textbf{Response of cryptocurrencies to MNA about inflation}} 
  \label{} 
    \begin{flushleft}
    {\medskip\small
 The table documents response of cryptocurrency prices to inflation MNA surprises. Each column reports estimates from the regression $R_t = a + b Surprise_t + \epsilon_t.$ for different cryptocurrency. Cryptocurrency returns are measured over 30-minutes window around MNA.}
    \medskip
    \end{flushleft}
\begin{tabular}{@{\extracolsep{0pt}}lcccccc} 
\\[-1.8ex]\hline 
\hline \\[-1.8ex] 
 & \multicolumn{6}{c}{\textit{Dependent variable:}} \\ 
\cline{2-7} 
\\[-1.8ex] & \multicolumn{6}{c}{Return} \\ 
\\[-1.8ex] & (1) & (2) & (3) & (4) & (5) & (6)\\ 
\hline \\[-1.8ex] 
 Surprise & -0.238$^{**}$ & -0.014 & -0.114 & -0.176$^{**}$ & -0.038 & -0.029 \\ 
  & [-2.440] & [-0.144] & [-0.972] & [-2.218] & [-0.155] & [-0.181] \\ 
  & & & & & & \\ 
 Constant & 0.051 & -0.091 & -0.141 & 0.100 & 0.214 & 0.031 \\ 
  & [0.642] & [-1.073] & [-1.329] & [1.447] & [0.986] & [0.213] \\ 
  & & & & & & \\ 
\hline \\[-1.8ex] 
Coin/token & BTC & ETH & XRP & BCH & LTC & ADA \\ 
Observations & 195 & 125 & 103 & 91 & 105 & 77 \\ 
Adjusted R$^{2}$ & 0.025 & -0.008 & -0.001 & 0.042 & -0.009 & -0.013 \\ 
\hline 
\hline \\[-1.8ex] 
\textit{Note:}  & \multicolumn{6}{r}{$^{*}$p$<$0.1; $^{**}$p$<$0.05; $^{***}$p$<$0.01} \\ 
\end{tabular} 
\end{table}

\begin{table}[!htbp] \centering 
  \caption{\textbf{Response of Bitcoin to MNA about growth}} 
  \label{} 
      \begin{flushleft}
    {\medskip\small
 The table documents response of cryptocurrency prices to growth MNA surprises. Each column reports estimates from the regression $R_t = a + b Surprise_t + \epsilon_t.$ for different cryptocurrency. Cryptocurrency returns are measured over 30-minutes window around MNA.}
    \medskip
    \end{flushleft}
\begin{tabular}{@{\extracolsep{-4pt}}lccccccc} 
\\[-1.8ex]\hline 
\hline \\[-1.8ex] 
 & \multicolumn{7}{c}{\textit{Dependent variable:}} \\ 
\cline{2-8} 
\\[-1.8ex] & \multicolumn{7}{c}{Bitcoin Return} \\ 
\\[-1.8ex] & (1) & (2) & (3) & (4) & (5) & (6) & (7)\\ 
\hline \\[-1.8ex] 
 Surprise & -0.179$^{**}$ & 0.004 & -0.001 & 0.008 & 0.050 & 0.001 & 0.014 \\ 
  & [-2.545] & [0.033] & [-0.024] & [0.174] & [0.848] & [0.009] & [0.286] \\ 
  & & & & & & & \\ 
 Constant & 0.025 & 0.009 & -0.042 & -0.045 & -0.010 & 0.020 & -0.012 \\ 
  & [0.194] & [0.067] & [-0.535] & [-0.566] & [-0.177] & [0.176] & [-0.210] \\ 
  & & & & & & & \\ 
\hline \\[-1.8ex] 
News & PCE & GDP & NFP & Unemployment & PMI & Retail & Construction \\ 
Observations & 33 & 33 & 98 & 98 & 98 & 97 & 97 \\ 
Adjusted R$^{2}$ & 0.146 & -0.032 & -0.010 & -0.010 & -0.003 & -0.011 & -0.010 \\ 
\hline 
\hline \\[-1.8ex] 
\textit{Note:}  & \multicolumn{7}{r}{$^{*}$p$<$0.1; $^{**}$p$<$0.05; $^{***}$p$<$0.01} \\ 
\end{tabular} 
\end{table}

\begin{table}[!htbp] \centering 
  \caption{\textbf{Response of 5-year Treasury yield to MNA of two types}} 
  \label{}
        \begin{flushleft}
    {\medskip\small
The table documents relation between interest rate change and MNA surprise. Each column reports estimates from the regression $\Delta Rate_t = a + Surprise_t + \epsilon_t.$ for different type of MNA. Bitcoin returns as well as interest rate change are measured over 30-minutes window around MNA.}
    \medskip
    \end{flushleft}
\begin{tabular}{@{\extracolsep{0pt}}lcc} 
\\[-1.8ex]\hline 
\hline \\[-1.8ex] 
 & \multicolumn{2}{c}{\textit{Dependent variable:}} \\ 
\cline{2-3} 
\\[-1.8ex] & \multicolumn{2}{c}{$\Delta Rate$} \\ 
\\[-1.8ex] & (1) & (2)\\ 
\hline \\[-1.8ex] 
 Surprise & 0.004$^{***}$ & 0.010$^{***}$ \\ 
  & [10.103] & [7.510] \\ 
  & & \\ 
 Constant & 0.001$^{*}$ & 0.0003 \\ 
  & [1.712] & [0.350] \\ 
  & & \\ 
\hline \\[-1.8ex] 
News Type & Growth & Inflation \\ 
Observations & 2,060 & 594 \\ 
Adjusted R$^{2}$ & 0.047 & 0.085 \\ 
\hline 
\hline \\[-1.8ex] 
\textit{Note:}  & \multicolumn{2}{r}{$^{*}$p$<$0.1; $^{**}$p$<$0.05; $^{***}$p$<$0.01} \\ 
\end{tabular} 
\end{table}

\begin{table}[!htbp] \centering 
  \caption{\textbf{Relation between Bitcoin returns and change in 5-year Treasury yield during MNA of two types}} 
  \label{} 
          \begin{flushleft}
    {\medskip\small
 The table documents relation between Bitcoin returns and interest rate change during MNA. Each column reports estimates from the regression $R_t = a + \Delta Rate_t + \epsilon_t.$ for different type of MNA. Bitcoin returns as well as interest rate change are measured over 30-minutes window around MNA.}
    \medskip
    \end{flushleft}
\begin{tabular}{@{\extracolsep{0pt}}lcc} 
\\[-1.8ex]\hline 
\hline \\[-1.8ex] 
 & \multicolumn{2}{c}{\textit{Dependent variable:}} \\ 
\cline{2-3} 
\\[-1.8ex] & \multicolumn{2}{c}{Bitcoin return} \\ 
\\[-1.8ex] & (1) & (2)\\ 
\hline \\[-1.8ex] 
 $\Delta Rate$ & -0.033 & 2.032 \\ 
  & [-0.030] & [0.888] \\ 
  & & \\ 
 Constant & 0.006 & 0.012 \\ 
  & [0.170] & [0.212] \\ 
  & & \\ 
\hline \\[-1.8ex] 
News Type & Growth & Inflation \\ 
Observations & 2,060 & 594 \\ 
Adjusted R$^{2}$ & -0.0005 & -0.0004 \\ 
\hline 
\hline \\[-1.8ex] 
\textit{Note:}  & \multicolumn{2}{r}{$^{*}$p$<$0.1; $^{**}$p$<$0.05; $^{***}$p$<$0.01} \\ 
\end{tabular} 
\end{table}

\end{document}